\documentclass[twocolumn]{svjour3}          
\smartqed  
\usepackage{mathptmx}      
\usepackage{latexsym,amssymb,amsmath,graphicx,epsfig,cite,bbm,float,epsf,graphics}

\usepackage{epstopdf}
\usepackage[algo2e]{algorithm2e}
\usepackage{setspace}
\usepackage{bm}

\newcommand{\Real}{{\mathbbm{R}}}
\newcommand{\sbt}{\left\{b(t) \right\}_{t \geq 1}}

\newcommand{\srt}{\left\{r(t) \right\}_{t \geq 1}}

\newcommand{\vx}{\vec{x}}

\newcommand{\vr}{\vec{r}}
\newcommand{\ur}{\breve{\vec{r}}}
\newcommand{\uw}{\breve{\vec{w}}}

\newcommand{\vw}{\vec{w}}
\newcommand{\vf}{\vec{f}}

\newcommand{\mA}{\vec{A}}
\newcommand{\mI}{\vec{I}}

\newcommand{\mSig}{\bm{\Sigma}} 
\newcommand{\malpha}{\bm{\alpha}}

\newcommand{\nn}{\nonumber}
\newcommand{\Tr}{\mathrm{Tr}}

\newcommand{\vtw}{\tilde{\vec{w}}}
\newcommand{\mR}{\vec{R}}

\journalname{Signal, Image and Video Processing}
\begin{document}

\title{A new robust adaptive algorithm for underwater acoustic channel equalization
}

\titlerunning{A robust adaptive equalizer for underwater communications}        

\author{Dariush Kari         \and
        Muhammed Omer Sayin \and
        Suleyman Serdar Kozat, {\em Senior Member, IEEE}
}


\institute{D. Kari \at
              Department of Electrical and Electronics Engineering, Bilkent University, Bilkent, Ankara 06800, Turkey \\
              \email{kari@ee.bilkent.edu.tr}
           \and
           M. O. Sayin \at
              Coordinated Science Laboratory, University of Illinois at Urbana-Champaign, Urbana, IL 61801 USA \\
              \email{sayin2@illinois.edu}           
           \and
           S. S. Kozat \at
              Department of Electrical and Electronics Engineering, Bilkent University, Bilkent, Ankara 06800, Turkey \\
              \email{kozat@ee.bilkent.edu.tr}           
}

\date{Received: date / Accepted: date}

\maketitle

\begin{abstract}
  We introduce a novel family of adaptive robust equalizers for highly challenging underwater acoustic (UWA) channel equalization. Since the underwater environment is highly non-stationary and subjected to impulsive noise, we use adaptive filtering techniques based on a relative logarithmic cost function inspired by the ``competitive methods'' from the online learning literature. To improve the convergence performance of the conventional linear equalization methods, while mitigating the stability issues, we intrinsically combine different norms of the error in the cost function, using logarithmic functions. Hence, we achieve a comparable convergence performance to least mean fourth (LMF) equalizer, while significantly enhancing the stability performance in such an adverse communication medium. We demonstrate the performance of our algorithms through highly realistic experiments performed on accurately simulated underwater acoustic channels.
\end{abstract}

\section{Introduction}
\label{intro}
Underwater acoustic (UWA) communication has attracted much attention in recent years due to proliferation of new and exciting applications such as marine environmental monitoring and sea bottom resource exploitation\cite{Tao,mag}. However, due to the constant movement of waves, multi-path propagation, large delay spreads, Doppler effects, and frequency dependent propagation loss \cite{CasasantaG12}, the underwater acoustic channel is considered as one of the most detrimental communication mediums in use today\cite{sto4,state_art}. In these mediums channel equalization\cite{Vasudevan2007,Pinchas2011,proakis_book} plays a key role in providing reliable high data rate communication\cite{mag}. Furthermore, due to rapidly changing and unpredictable nature of underwater environment, such processing should be adaptive \cite{mag,Pinchas13,FernandesFM07}. However, the impulsive nature of the ambient noise in UWA channels, introduces stability issues for adaptive equalization \cite{DSSS}. To this end, in this paper we propose a new adaptive linear channel equalizer, which provides a relatively fast convergence rate as well as strong stability against the ambient noise in the UWA channels.\par
Although the additive white Gaussian noise (AWGN) model is widely used in digital and wireless communication contexts, this model is insufficient to appropriately address the ambient noise in UWA channels\cite{Sharbari1,Sharbari2,Sharbari3}. For example, in warm shallow waters, the high frequency noise component is dominated by the ``impulsive noise''\cite{local-imp,DjurovicL07,Li15} resulted from numerous noise sources such as marine life, shipping traffic, underwater explosives, and offshore oil exploration-production\cite{Hildebrand}. The impulsive noise consists of relatively short duration, infrequent, high amplitude noise pulses. In this paper, in order to rectify the undesirable effects of UWA channels, especially to mitigate the effects of the impulsive noise, we introduce a radical approach to adaptive channel equalization and seek to provide adaptive algorithms that show a fast convergence rate as well as robustness against the impulsive noise.\par
Linear adaptive channel equalizers (e.g., sign algorithm (SA), least mean squares (LMS) or least mean fourth (LMF) algorithms) are the simplest low complexity equalization methods. However, the conventional linear equalizers either have a slow convergence speed (e.g., sign algorithm (SA)), or suffer from stability issues (e.g., LMF) in impulsive noise environments, hence cannot fully address the problem \cite{omer}. These methods commonly use different powers (norms) of the error amount (the difference between the transmitted symbol and the estimate from the equalizer) as the cost function. As an example, the well-known sign algorithm (SA) uses {\small $C(e(t)) \triangleq E\left[|e(t)|\right]$} as the cost, while LMF and its family use even powers of error, i.e., $C(e(t)) \triangleq E\left[e(t)^{2n}\right]$, where $E\left[.\right]$ indicates the expectation. Then, an algorithm is derived to adaptively adjust the linear equalizer coefficients, based on minimization of the cost function using a stochastic gradient descent method \cite{omer,sayed_book}. However, there is always a trade-off between the convergence speed and the ``robustness'' of such algorithms \cite{omer}. We emphasize that in this context we define the robustness as insensitivity against the impulsive noise. In this sense, the algorithms that use lower powers of the error as the cost function (e.g., SA \cite{sayed_book}) provide a better robustness rather than the algorithms that use higher powers of the error. However, the algorithms based on higher powers of the error, usually exhibit faster convergence \cite{omer}.\par
The mixed norm algorithms, combine different norms of the error in the cost function, in order to achieve a better trade-off between robustness and convergence speed \cite{chambers1994,chambers1997}. Nevertheless, optimization of the combination parameters in such algorithms needs ``a priori'' knowledge of the noise and input signal statistics \cite{omer}. On the contrary, the mixture of experts algorithms \cite{garcia2005,garcia2006,silva2008}, adaptively learn the best combination parameters. However, such algorithms are infeasible in UWA scenarios, due to the high computational complexity resulted from running several different algorithms in parallel \cite{omer,kozat2010}.\par
It is important to note that the samples contaminated with noise impulses, contain little useful information \cite{kim1995} for updating the equalizer coefficients. Hence, it is sufficient for the robust algorithms to be less sensitive only against large error values, i.e., for small error values the algorithm can be as sensitive as the conventional least squares algorithms \cite{omer}.\par
In this paper, in order to obtain a comparable performance to the robust algorithms, while retaining the fast convergence of conventional least square methods, we use a logarithmic function as a regularization term in the cost function of the well-known adaptive methods. In this sense, we choose a conventional method that uses a power of the error as the cost, e.g, least mean squares (LMS), and improve that method through adding a logarithmic term to its cost function. Due to the characteristics of the logarithmic functions, when the error is high, e.g., when there is an impulse, the cost function resembles the cost function of the original method, while for the small errors a correction term is added. The correction term includes the higher powers of the error, which yields a faster convergence. Hence we intrinsically mitigate the effect of the impulsive noise pulses and provide an improved robustness, while increasing the convergence speed when there is no impulsive noise. Specifically we present two equalization algorithms, logarithmic cost least mean absolute (LCLMA) and logarithmic cost least mean square (LCLMS), based on the first and second powers of the error as in SA and LMS, and show the improved performance of our algorithms through highly realistic simulations.\par
The paper is organized as follows: In Section \ref{sec:prob} we introduce the notation and describe the problem mathematically. Then, in Section \ref{sec:log} we provide a family of stable and fast converging equalizers based on logarithmic cost functions, and we provide the performance analysis for our method in Section \ref{sec:performance}. We demonstrate the performance of the presented methods through highly realistic simulated channels in Section \ref{sec:sim} and conclude the paper with Section \ref{sec:con}.

\section{Problem description}\label{sec:prob}

In this paper, all vectors are column vectors and denoted by boldface lower case letters. For a vector $\vx$, $\vx^T$ is the ordinary transpose. We denote by $E[x]$ the expectation of the random variable $x$, and by Tr($\vec{A}$) the trace of the matrix $\vec{A}$. Furthermore, for a vector $\vw$ the weighted squared norm by a positive definite matrix $\mSig$ is defined as $\|\vw\|_{\mSig}^2 \triangleq \vw^T \mSig \vw$.\par
\begin{figure}
	\includegraphics[width=0.4\textwidth]{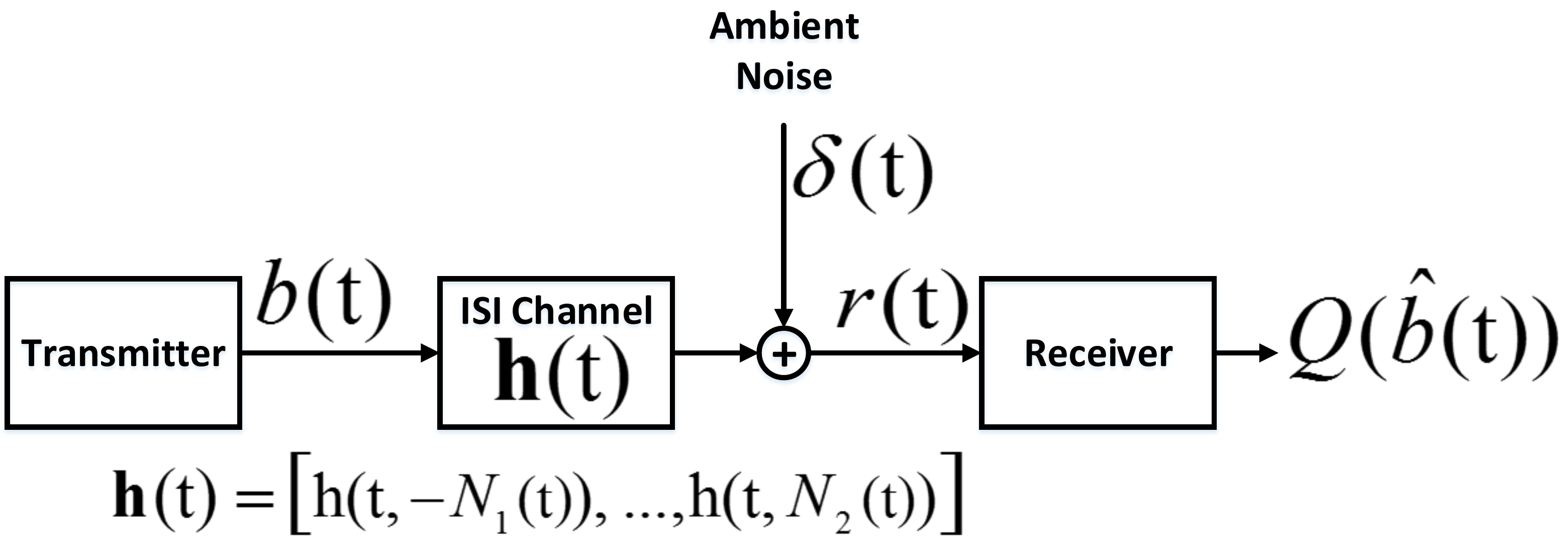}
	\caption{The block diagram of the model we use for the transmitted and received signals. The transmitted data $\sbt$ are BPSK modulated, then, pass through a time varying intersymbol interference (ISI) channel $\vec{h}(t)$. The received signal is the output of the ISI channel contaminated with the ambient noise $\delta(t)$. $Q(\hat{b}(t))$ is the hard estmation for the transmitted bit at time $t$.}
	\label{fig:TxRx}
\end{figure}
We denote the received signal by $\srt$, $r(t) \in \Real$, and our aim is to determine the transmitted bits $\sbt$, $b(t) \in \{-1,1\}$. As depicted in Fig. \ref{fig:TxRx}, using the baseband discrete time model (see Chapter 3 of \cite{goldsmith_book}) representation for the UWA channel, the received signal at time $t = 1,2, \dots$ is
\[
r(t) = \sum_{\tau=-N_1(t)}^{N_2(t)} b(t-\tau) h(t,\tau) + \delta(t),
\]
where $b(t)$ is the transmitted bit at time $t$. $h(t,\tau)$ is the overall impulse response of the channel at time $t$, $N_1(t)$ and $N_2(t)$ respectively represent the lengths of anti-causal and causal parts of the channel response at time $t$, and $\delta(t)$ represents the ambient noise of the channel, which is represented as
\[
\delta(t)=\delta_g(t)+\delta_i(t),
\]
where $\delta_g(t)$ indicates the white Gaussian part of the noise and $\delta_i(t)$ is the impulsive part. Note that the effects of time delay and phase deviations are usually addressed at the front-end of the receiver, hence, in this paper we do not deal with these problems explicitly.\par
\begin{figure}
	\includegraphics[width=0.4\textwidth]{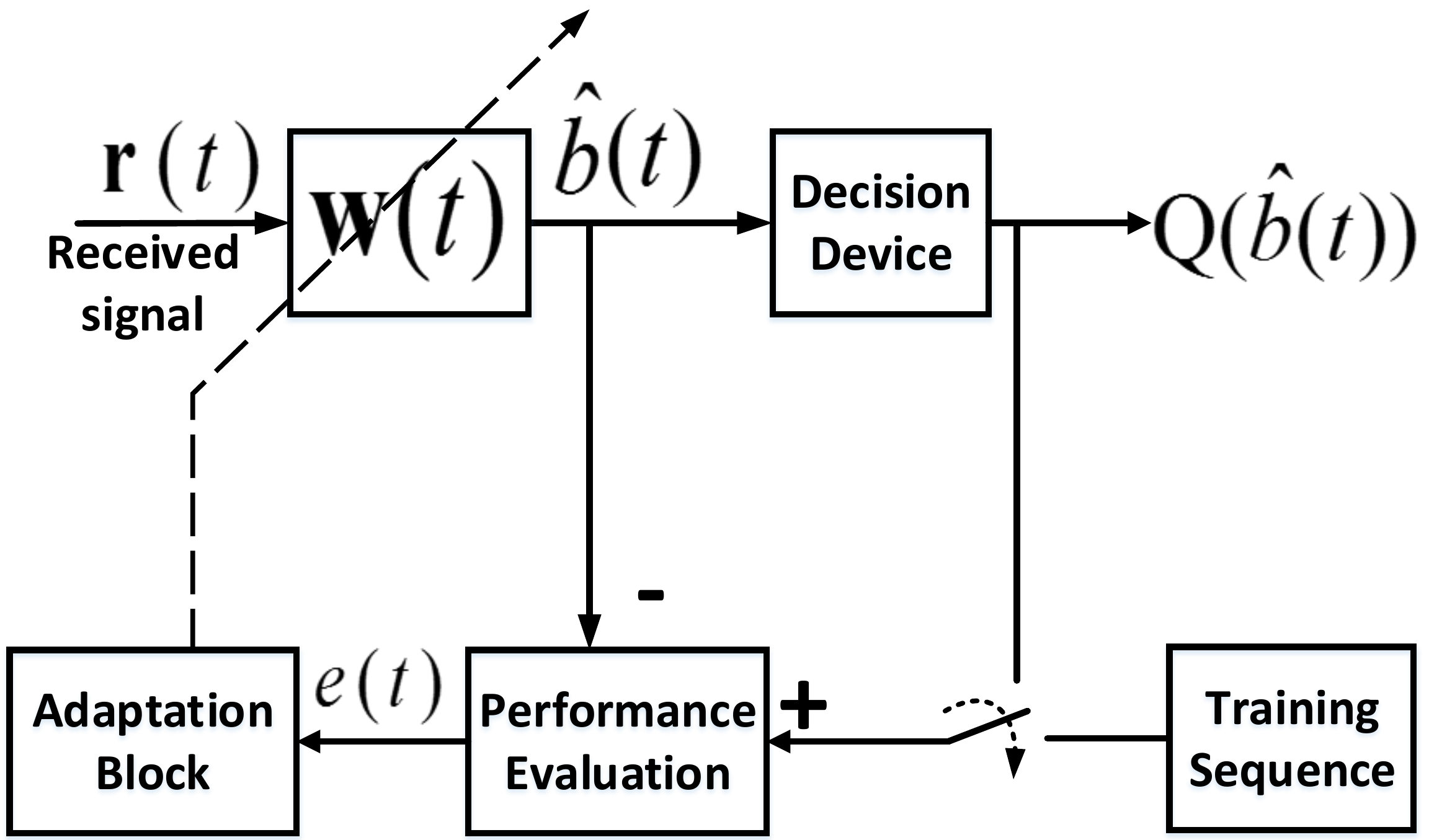}
	\caption{The block diagram of an adaptive equalizer. In this paper, we present new algorithms for the adaptation block (which determines how should $\vw$ be updated based on the error amount), which yields improved performance over the conventional methods.}
	\label{adp_eq}
\end{figure}

Our aim is to estimate the transmitted bits $\sbt$ according to the channel outputs $\srt$. To obtain the estimate $\hat{b}(t)$ we use a linear channel equalizer as depicted in Fig. \ref{adp_eq}, which is mathematically represented as $ \hat{b}(t) = \vw^T(t) \vr(t)$, where $\vr(t) \triangleq [r(t+L_a), \dots, r(t-L_c)]^T$, $L_a$ and $L_c$ are the length of anti-causal and causal parts of the equalizer, respectively, and $\vw(t) \triangleq [w_{-L_a}(t), \dots, w_{L_c}(t)]^T$ is the tap weights vector of the linear equalizer at time $t$. We define the error in estimating the transmitted bit $b(t)$ as $e(t) = b(t) - \hat{b}(t)$.\par
In order to update the coefficients of the adaptive equalizer, i.e., the tap weights vector $\vw(t)$, a cost function $C(e(t))$ is defined, e.g., in LMS method the cost is defined as $C(e(t))=E[e(t)^2]$. Then, we derive an update algorithm for $\vw(t)$ based on minimization of this cost function using the stochastic gradient descent method \cite{sayed_book}. Therefore, we update the tap weights vector as follows
\[{\bf{w}}(t + 1) = {\bf{w}}(t) - \frac{1}{2}\mu {\nabla _{{\bf{w}}(t)}}C(e(t)),\]
where $\nabla _{\vw(t)}C(e(t))$ denotes an instantaneous approximation of the gradient of the cost function $C(e(t))$ with respect to $\vw(t)$, i.e., the gradient obtained by removing the expectation and taking the gradient of the term inside that \cite{sayed_book}. As an example, in LMS method $C(e(t))=E[e(t)^2]$, hence we update the tap weights vector of an LMS equalizer as
\begin{equation}\label{eq:lms_up}
\vw(t + 1) = \vw(t) + \mu e(t) \vr(t),
\end{equation}
where $\mu>0$ is the learning rate.\par
The updating expression in \eqref{eq:lms_up} is the well-known LMS update. However, we seek to provide a more robust updating algorithm through minimization of a different cost function.
Note that the cost functions of the form {\small $ C(e(t))=E[|e(t)|^k]$}, which consist of only the $k$th power of the error, either have a slow convergence, or do not perform well, from the stability viewpoint, in an impulsive noise environment \cite{omer}. Therefore, in Section \ref{sec:log} we use a logarithmic term in the cost function, which intrinsically introduces different powers of the error into the cost function. As a result, when an impulsive error occurs, the algorithm mitigates the effect of that sample in updating the equalizer coefficients by simply using the lower order norms of the error, whereas in impulse-free environments the algorithm accelerates the convergence using higher order norms of the error.\par
\section{Adaptive equalizers based on logarithmic cost functions}\label{sec:log}
In this section we explain our method mathematically and especially introduce two new channel equalizers based on the logarithmic functions. First we describe the most important characteristics of the logarithmic functions in this context.\par
Suppose that $\Phi(e(t)) : \Real \to [0,\infty)$ represents a non-negative function of the error, which is an increasing function of $|e(t)|$. It can be straightforwardly shown that the function $L(e(t)) \triangleq \ln (1+\Phi(e(t)))$ has the following properties:
\begin{enumerate}
	\item $ \mathop {\lim }\limits_{|e(t)| \to \infty } \frac{{L(e(t))}}{{\Phi (e(t))}} = 0. $
	\item By Maclaurin series of the natural logarithm:\\ 
	As $\Phi(e(t)) \to 0 : L(e(t)) \approxeq \sum\limits_{i = 1}^\infty  {\frac{{{{( - 1)}^{i + 1}}\Phi {{(e(t))}^i}}}{{i}}} $
\end{enumerate}
The first property reveals that for high error values, e.g., noise impulses, the logarithmic function $L(e(t))$ is negligible relative to the original cost function $\Phi(e(t))$. However, the second property shows how the greater powers of the cost function $\Phi(e(t))$ are contributed to the logarithmic function $L(e(t))$ when the error values are small.\par
By leveraging these two properties, we proceed to define a new cost function based on a primary cost function $\Phi(.)$. Thus, we define the new cost as
\[C(e(t)) \triangleq \Phi(e(t))-\frac{1}{a} \ln (1+a\Phi(e(t))),\]
where $a>0$ is a design parameter with a typical value $a=1$. Regarding the properties of the logarithmic function discussed above, we deduce that when $a \Phi(e(t)) \ll 1$
\[C(e(t)) \to \frac{a}{2}\Phi^2(e(t))-\frac{a^2}{3}\Phi^3(e(t))+\dots\]
Moreover,
\[C(e(t)) \to \Phi(e(t)), \quad \textrm{as}\ \ e(t) \to \pm \infty.\]
The new cost function $C(e(t))$ is a convex function of $e(t)$ \cite{omer}. Therefore, in the new algorithm, we seek to minimize a cost function that is mainly consisted of the first and second powers of the primary cost function $\Phi(e(t))$, based on the error amount. For this, we use the stochastic gradient method \cite{sayed_book} to derive a recursion expression for updating $\vw(t)$. This yields,
\begin{align*}
{\bf{w}}(t + 1) & = {\bf{w}}(t) - \frac{1}{2}\mu {\nabla _{{\bf{w}}(t)}}C(e(t)) \\
& = \vw(t) - \mu\ \dfrac{a\Phi(e(t))}{1+a\Phi(e(t))}\left[\nabla_{\vw(t)} \Phi(e(t))\right].
\end{align*}
Note that in this framework we adopt a well-known cost function, e.g., $\Phi(e(t))=E[e^2(t)]$, as the primary cost function $\Phi(e(t))$. However, the objective functions $\Phi^2(e(t))$, e.g., $E[e^2(t)]^2$, and $\Phi(e(t)^2)$, e.g., $E[e^4(t)]$ yield the same stochastic gradient updates after removing the expectation in this paper. Suppose that $\Phi(e(t))$ is the expectation of another function $\varphi(e(t))$, i.e., $\Phi(e(t))=E[\varphi(e(t))]$. Then, using the instantaneous approximation for $\Phi(e(t))$ \cite{sayed_book}, the general stochastic gradient update is given by
\[\vw(t+1)= \vw(t) + \mu\ \dfrac{a\varphi(e(t))}{1+a\varphi(e(t))}\left[\nabla_{e(t)} \varphi(e(t))\right] \vr(t).\]
\par
Here we present two equalization methods based on the introduced approach. We then show the superior performance of these methods through highly realistic experiments in Section \ref{sec:sim}.
\begin{enumerate}
\item {\em The logarithmic cost least mean square (LCLMS)}:
Here we adopt $\varphi(e(t))=e^2(t)$, which yields the following update on the tap weights vector
\begin{align*}
	\vw(t+1) & = \vw(t) + \mu\ \dfrac{a\ e^2(t)}{1+a\ e^2(t)}\left[2 e(t)\right] \vr(t) \\
	         & = \vw(t) + \mu'\ \dfrac{a\ e^3(t)}{1+a\ e^2(t)} \vr(t).
\end{align*}

\item {\em The logarithmic cost least mean absolute (LCLMA)}:
In this case, we adopt $\varphi(e(t))=|e(t)|$, and we obtain the following update on the tap weights vector
\begin{align*}
\vw(t+1) & = \vw(t) + \mu\ \dfrac{a\ |e(t)|}{1+a\ |e(t)|}\left[sign(e(t))\right] \vr(t) \\
& = \vw(t) + \mu\ \dfrac{a\ e(t)}{1+a\ |e(t)|} \vr(t).
\end{align*}
\end{enumerate}
As shown in the simulations, the LCLMS algorithm results in an improved convergence speed over the conventional LMS algorithm, while achieving the comparable stability to LMS method. Similarly, we can achieve an improved convergence speed performance over the conventional SA algorithm by using LCLMA algorithm, while preserving the robustness against impulsive noise. Hence, the LCLMA is an elegant alternative to the conventional methods for UWA channel equalization.\par
{\em Remark 1:} These methods can be directly applied to decision feedback equalizers (DFEs). In this sense, we can update the feed-forward and feedback filter coefficients in such equalizers, with different methods. However, since we use the same error as the feed-forward filter, to update the feedback filter, this part is also affected by the impulsive noise. Thus, we use the same robust adaptive algorithm in both feed-forward and feedback parts of the equalizer. To this end, we append the past decided symbols to the received signal vector as
\[\ur(t) \triangleq [r(t), \dots, r(t-h+1),\bar{b}(t-1), \dots, \bar{b}(t-h_f)]^T,\]
where $h_f$ is the length of the feedback part of the equalizer. Also, $\bar{b}(t)=Q(\hat{b}(t))$ denotes the hard decided (quantized) symbol for the transmitted bit $b(t)$. Furthermore, corresponding to this extension in the received signal vector, we merge the feed-forward and feedback equalizers to obtain an extended filter of length $h+h_f$ as $\uw(t) \triangleq [\vw^T(t) \quad \vf^T(t)]^T$, where $\vf(t)$ represents the feedback filter at time $t$. Hence, $\hat{b}(t) = \uw^T(t)\ \ur(t)$. We update the extended tap weights vector $\uw(t)$ with the desired method, e.g., LCLMA method.\par
{\em Remark 2:} For further robustness, we normalize the updates with respect to the received signal vector to make the updates more independent from the received signal. Hence, we proceed to minimize $C(\frac{e(t)}{\|\vr(t)\|})$ with respect to $\vw$, which yields the following update
\[
\vw(t+1) = \small{\vw(t) + \mu\ \dfrac{a\varphi(\frac{e(t)}{\|\vr(t)\|})}{1+a\varphi(\frac{e(t)}{\|\vr(t)\|})}\left[\nabla_{\frac{e(t)}{\|\vr(t)\|}} \varphi(\frac{e(t)}{\|\vr(t)\|})\right] \frac{\vr(t)}{\|\vr(t)\|}}.
\]
As an example, the normalized LCLMA will be
\begin{align*}
\vw(t+1) & = \vw(t) + \mu\ \dfrac{a\ \frac{e(t)}{\|\vr(t)\|}}{1+a\ |\frac{e(t)}{\|\vr(t)\|}|} \frac{\vr(t)}{\|\vr(t)\|}\\
& = \vw(t) + \mu\ \dfrac{a\ e(t)}{\|\vr(t)\|(\|\vr(t)\|+a\ |e(t)|)} \vr(t).
\end{align*}
\section{Performance analysis}\label{sec:performance}
In order to analyze the tracking performance of the introduced algorithms, we assume a random walk model \cite{sayed_book} for the tap weights vector $\vw^*(t)$ that yields the minimum mean squared error such that
\begin{equation}\label{eq:random}
\vw^*(t+1) =\vw^*(t)+\malpha(t).
\end{equation}
We define $\vtw(t) \triangleq \vw^*(t) - \vw(t)$ and $\malpha(t) \in \Real^h$ is a zero-mean vector process with covariance matrix $E[\malpha(t)\malpha(t)^T] = \mA$. Assuming that $\malpha(t)$ is independent from the received and noise signals, we have the following general weighted-energy recursion for an adaptive filter with error nonlinearity
\begin{align}\label{eq:weighted}
E\left[\|\vtw(t+1)\|_{\mSig}^2\right] = &\,E\left[\|\vtw(t)\|_{\mSig}^2\right] - \mu 2 E\left[e_{a}^{\mSig}(t)g(e(t))\right]\nn\\
&+\mu^2E\left[\|\vr(t)\|_{\mSig}^2g^2(e(t))\right] + \Tr(\mSig\mA),
\end{align}
where $e_{a}^{\mSig}(t)\triangleq \vr^T(t)\mSig\vtw(t)$ is the weighted {\em a priori} estimation error and $g(e(t))$ is the nonlinear error function \cite{tareq2003,omer}. Moreover, $e(t)=e_a(t)+n(t)$, in which $e_a(t)$ is a priori estimation error where $\mSig=\mI$, and $n(t)$ indicates the estimation noise, i.e., the error resulted from the optimal linear estimator. For the proposed algorithms, $g(e(t))$ is defined as
\begin{equation}\label{eq:g_func}
g(e(t)) \triangleq \frac{\partial \varphi(e(t))}{\partial e(t)}\frac{a \varphi(e(t))}{1+a \varphi(e(t))}.
\end{equation}

In our analysis, we use the following assumptions:
\begin{description}
	\item[{\it Assumption 1:}] \hfill \\
	The noise signal $n(t)$ is a zero-mean independently and identically distributed (i.i.d.) Gaussian random variable and independent from $\vr(t)$. The received signal $\vr(t)$ is also a zero-mean i.i.d. Gaussian random variable with the auto-correlation matrix $\mR \triangleq E\left[\vr(t)\vr^T(t)\right]$.
	\item[{\it Assumption 2:}] \hfill \\
	The a priori estimation error $e_{a}(t)$ has Gaussian distribution and is jointly Gaussian with the weighted a priori estimation error $e_{a}^{\mSig}(t)$ for a positive definite symmetric matrix $\mSig$. This assumption is reasonable by the Assumption 1, whenever $h$ is sufficiently large and the learning rate $\mu$ is sufficiently small \cite{tareq2003}.
	\item[{\it Assumption 3:}] \hfill \\
	The random processes $\|\vr(t)\|_{\mSig}^2$ and $g^2(e(t))$ are uncorrelated, which results the following separation
	\begin{align*}
	E\left[\|\vr(t)\|_{\mSig}^2g^2(e(t))\right] = E\left[\|\vr(t)\|_{\mSig}^2\right]E\left[g^2(e(t))\right].
	\end{align*}
\end{description}

Assumptions 1-3 and the Price's Theorem \cite{sayed_book} lead the weighted-error recursion \eqref{eq:weighted} to
\begin{align*}
E\left[\|\vtw(t+1)\|_{\mSig}^2\right] = \,&E\left[\|\vtw(t)\|_{\mSig}^2\right] -\mu2h_G\left(e(t)\right)E\left[\|\vtw(t)\|_{\mSig\mR}^2\right]\nn\\
&+ \mu^2E\left[\|\vr(t)\|_{\mSig}^2\right]h_U\left(e(t)\right)  + \Tr(\mSig\mA),
\end{align*}
where
\begin{align*}
h_G(e(t))\triangleq \frac{E[e(t)g(e(t))]}{E[e^2(t)]}, \;\; h_U(e(t))\triangleq E\left[g^2(e(t))\right].
\end{align*}
Therefore, at the steady-state, i.e., $\lim\limits_{t \to \infty}E\left[\|\vtw(t+1)\|_{\mSig}^2\right]=\lim\limits_{t \to \infty}E\left[\|\vtw(t)\|_{\mSig}^2\right]$ , we have
\[
2\mu h_G(e(t))E\left[\|\vtw(t)\|_{\mSig\mR}^2\right] = \mu^2 E\left[\|\vr(t)\|_{\mSig}^2\right]h_U(e(t)) + \Tr(\mSig\mA).
\]
As a result of \cite{omer} and the previous assumptions, we arrive at the following expression for the tracking excess mean squared error (EMSE) \cite{sayed_book} in LCLMA method, which is shown by $\eta$, when the learning rate $\mu$ is sufficiently small.
\begin{align}\label{eq:track}
\eta_{\mathrm{LCLMA}} = \frac{\mu a \Tr(\mR)\sigma_n^2 + \mu^{-1} a^{-1}\Tr(\mA)}{2-\mu a \Tr(\mR)}.
\end{align}

However, although we assumed the white Gaussian noise model, in underwater acoustic communication the impulsive noise is a common problem \cite{local-imp,Hildebrand}. Therefore, we also provide the performance analysis in an impulsive noise environment scenario.\par
\noindent
{\bf Impulsive noise model:}
Since the received signal is subjected to impulsive noise, we model the estimation noise as a summation of two independent random terms~\cite{wang1997,chan2004} as
\begin{align*}
n(t) = v(t) + z(t)\gamma(t),
\end{align*}
where $v(t)$ is the ordinary AWGN noise signal that is zero-mean Gaussian with variance $\sigma_v^2$ and $\gamma(t)$ is the impulse-noise that is also zero-mean Gaussian with significantly large variance $\sigma_{\gamma}^2$. Here, $z(t)$ is generated through a Bernoulli random process and determines the occurrence of the impulses in the noise signal with $p_Z(z(t)=1) = \nu_i$ and $p_Z(z(t)=0)=1-\nu_i $ where $\nu_i$ is the frequency of the impulses. The overall probability density function of the noise signal $n(t)$ is given by
\begin{align*}
p_n(n(t)) = \frac{1-\nu_i}{\sqrt{2\pi}\sigma_{v}}\mathrm{exp}\left(-\frac{n^2(t)}{2\sigma_{v}^2}\right) + \frac{\nu_i}{\sqrt{2\pi}\sigma_{n}}\mathrm{exp}\left(-\frac{n^2(t)}{2\sigma_{n}^2}\right),
\end{align*}
where $\sigma_n^2 = \sigma_{v}^2 + \sigma_{\gamma}^2$.\\

We particularly provide the steady-state performance analysis of the LCLMA algorithm (for which $\varphi(e(t)) = |e(t)|$) in the impulsive noise environments, since we motivate the LCLMA algorithm as an improved alternative to the SA, in the sense of convergence speed. Hence, at the steady-state, \eqref{eq:weighted} yields
\begin{align}
E\left[\|\vr(t)\|_{\mSig}^2\right] = \frac{2 E\left[\frac{a\ e_a^{\mSig}(t)\ e(t)}{1+a|e(t)|}\right] - \mu^{-1}\Tr(\mSig\mA)}{\mu E\left[\frac{a^2e^2(t)}{\left(1+a|e(t)|\right)^2}\right]}.\label{eq:imp-steady}
\end{align}
By evaluating \eqref{eq:imp-steady} for $\mSig=\mI$ as in \cite{omer}, and by the fact that $\sigma_n^2 \gg \sigma^2_{e_a}$, we achieve the following tracking EMSE in an impulsive noise environment
\begin{align}
\eta_{\mathrm{LCLMA}}^{*} = \frac{\mu\Tr(\mR)\left(\nu_i + a^2(1-\nu_i)\sigma_{v}^2\right) + \mu^{-1}\Tr(\mA)}{a(1-\nu_i)(2-a\mu\Tr(\mR))+\sqrt{\frac{8}{\pi}}\frac{\nu_i}{\sigma_n}}.\label{eq:EMSE-IMP}
\end{align}
We extend these results to the LCLMA-DFE equalizer that was discussed in Section \ref{sec:log}. In BPSK modulation $E[b_i^2(t)]=1$, thus, it follows that $E[\bar{b}_i^2(t)]=1$. This yields
\[
\Tr(\mR_{DFE})= \Tr(\mR_{linear})+h_f,
\]
which results in the following expression for the tracking EMSE of a LCLMA-DFE in an impulsive noise scenario
\begin{small}
\begin{align}
\eta_{\mathrm{LCLMA-DFE}}^{*} = \frac{\mu(\Tr(\mR)+h_f)\left(\nu_i + a^2(1-\nu_i)\sigma_{v}^2\right) + \mu^{-1}\Tr(\mA)}{a(1-\nu_i)(2-a\mu(\Tr(\mR)+h_f))+\sqrt{\frac{8}{\pi}}\frac{\nu_i}{\sigma_n}},\label{eq:EMSE-IMP}
\end{align}
\end{small}
where, we have again assumed that a model similar to \eqref{eq:random} holds for the optimal tap weights vector, i.e., $\uw(t)$, in which $\malpha(t) \in \Real^{h+h_f}$.\par
As a result, we have the following expression for the MSE of an LCLMA equalizer in an impulsive noise environment
\begin{align*}
\lim\limits_{t \to \infty}E[|e(t)|^2] & = \lim\limits_{t \to \infty}E[|e_a(t)|^2]+E[|n(t)|^2]\\
& = \eta + \nu_i (\sigma_n^2) + (1-\nu_i) (\sigma_\nu^2),
\end{align*}
where $\eta$ is either $\eta_{LCLMA}$ or $\eta_{LCLMA-DFE}$, regarding which equalizer is used.
\section{Simulation results}\label{sec:sim}
We examine the performance of our algorithm under a highly realistic underwater acoustic channel equalization scenario through a highly accurate modeling of the real life channels introduced in \cite{Qarabaqi}. Particularly, we set the surface height to 100m, place the transmitter and receiver antennas at the height of 20m and 50m, and the channel distance is 1000m. We evaluate the performance of the SA, LMS and the introduced LCLMA and LCLMS equalization algorithms, both in DFE and linear equalizers. All of the results are averaged over 10 repetitions. In all of the experiments, we have sent 10000 repeated Turyn sequences of 28 bits over the decimated underwater acoustic channel \cite{turyn}. We compare normalized time accumulated squared errors (NASE) of the algorithms, which we define as $NASE \triangleq \dfrac{\sum_{t=1}^{n}(\hat{b}(t)-b(t))^2}{n}$.\par
Furthermore, we have used a mixture of Gaussian and 10\% impulsive noise model, i.e., $\nu_i=0.1$, and the variance of the impulsive noise is set to $10^4$ times that of the Gaussian noise. We have used feed-forward filters of length 362, and feedback filters of length 150 (in DFE equalizers), with the learning rates set to $\mu = 0.1$ for feed-forward filters and $\mu = 0.01$ for feedback filters.\par
\begin{figure}
	\includegraphics[width=3.5 in]{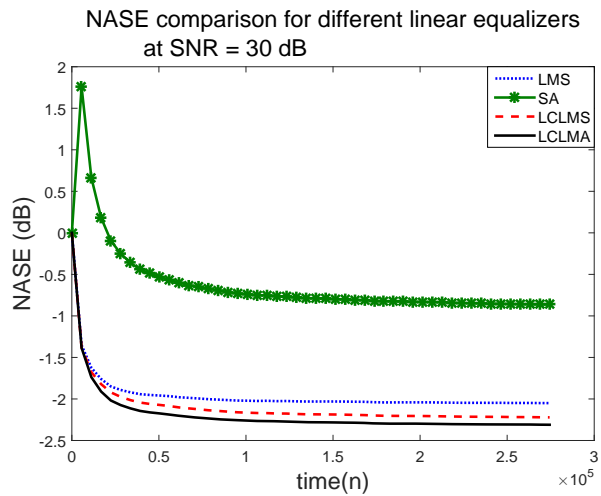}
	\caption{NASE of different linear equalizers, in a 10\% impulsive noise environment. This figure shows the superior convergence performance of the LCLMA method at SNR = 30 dB.}
	\label{fig:mse_linear}
\end{figure}
\begin{figure}
	\includegraphics[width=3.5 in]{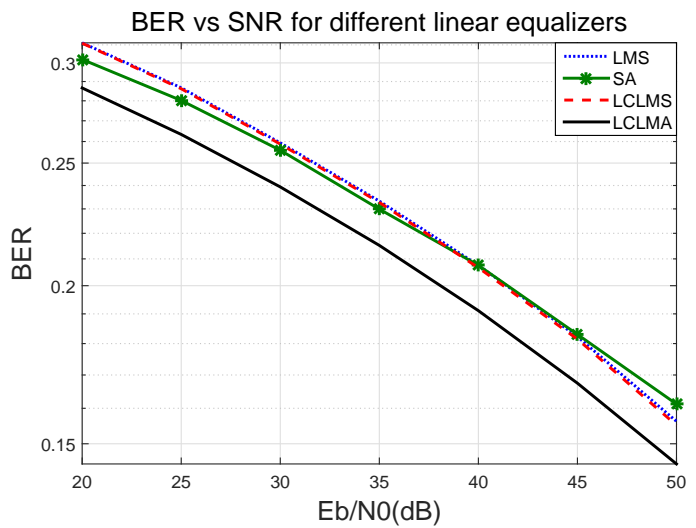}
	\caption{Bit error rate of different linear equalizers in different SNRs, in a 10\% impulsive noise environment. This figure shows the superior BER performance of the LCLMA method in all SNR values.}
	\label{fig:ber_linear}
\end{figure}

In the first experiment we compare the performance of different linear equalizers, all in training mode, i.e., the transmitted bits are known by the receiver for computing the soft error. The normalized accumulated squared error in Fig. \ref{fig:mse_linear} shows the superior performance of the proposed LCLMA algorithm at $\frac{E_b}{N_0}=30 dB$, where the variance of the Gaussian noise is $\frac{N_0}{2}$. In addition the LCLMS algorithm achieves the performance of the LMS algorithm, while the SA method outperforms LMS and LCLMS methods in lower SNRs. However, as depicted in Fig. \ref{fig:ber_linear} in high SNRs, the LMS and LCLMS methods deliver a better BER performance than SA, since the effect of the impulsive noise diminishes in higher SNRs. Therefore, in high SNRs the less robust but faster converging methods such as LMS or LCLMS methods, outperform the SA equalizer. Nevertheless, in all SNR values, the introduced LCLMA algorithm shows a better BER performance, which shows the superior performance, both in robustness as well as the convergence speed, of the LCLMA algorithm.\par

\begin{figure}
	\includegraphics[width=3.5 in]{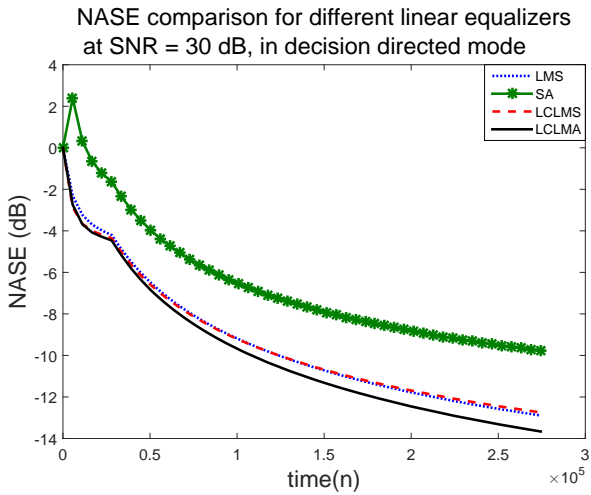}
	\caption{NASE of linear equalizers in decision directed mode with 10\% training data, in a 10\% impulsive noise environment.}
	\label{fig:mse_linear_dd}
\end{figure}

\begin{figure}
	\includegraphics[width=3.5 in]{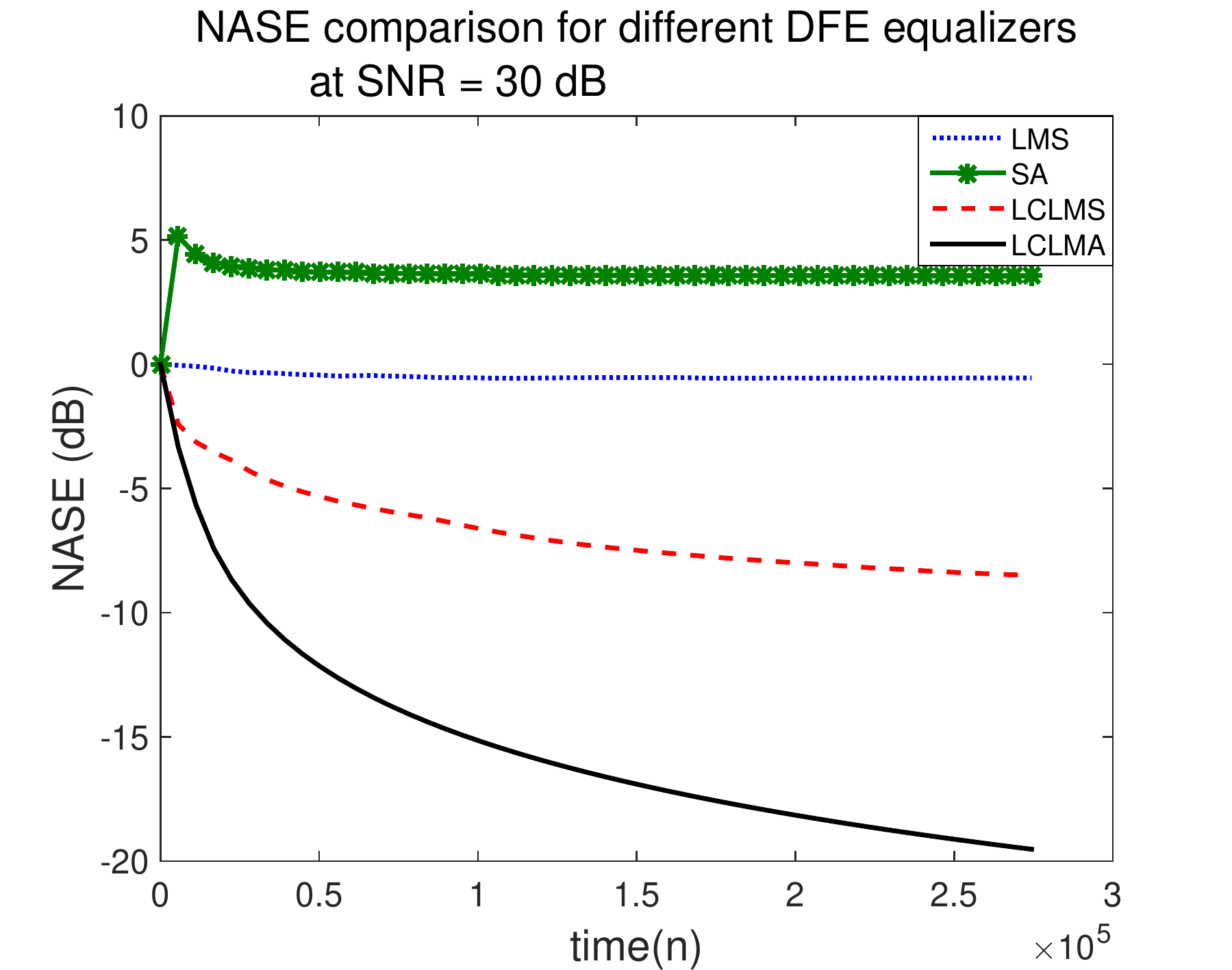}
	\caption{NASE of diffrent decision feedback equalizers (DFE), in a 10\% impulsive noise environment. This figure shows the superior performance of the LCLMA method.}
	\label{fig:mse_dfe}
\end{figure}

\begin{figure}
	\includegraphics[width=3.5 in]{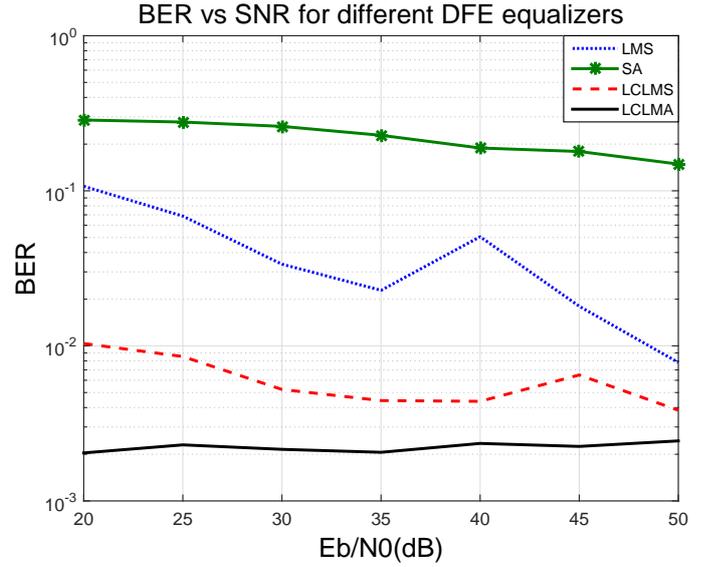}
	\caption{Bit error rate of different decision feedback equalizers (DFE) in different SNRs, in a 10\% impulsive noise environment. This figure shows the superior performance of the LCLMA method in all SNR values.}
	\label{fig:ber_dfe}
\end{figure}

We have also provided the performance of the algorithms in decision directed mode, with 10\% of the data used for training. As depicted in Fig. \ref{fig:mse_linear_dd}, also in this case the LCLMA algorithm delivers a better performance than other methods.
In the third experiment, we have implemented the algorithms in a decision feedback framework. We have used the same algorithm for feed-forward and feedback parts of each equalizer. The learning rate of the feedback part is $\mu=0.01$ in all of the equalizers. As depicted in Fig. \ref{fig:mse_dfe} and Fig. \ref{fig:ber_dfe} the LCLMA and LCLMS algorithms show a highly superior performance relative to the conventional SA and LMS methods in a DFE equalizer.

\begin{figure}
	\includegraphics[width=3.5 in]{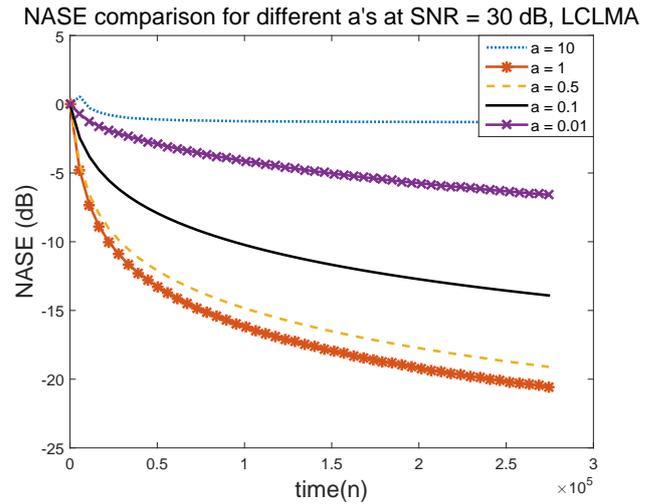}
	\caption{NASE of LCLMA linear equalizer with different $a$ values, in a 10\% impulsive noise environment.}
	\label{fig:nase_a}
\end{figure}

In the last experiment, we have investigated the effect of the design parameter $a$ on the performance of a linear equalizer. The results in Fig. \ref{fig:nase_a} reveal that for $\nu_i=0.1$ at SNR=30dB, $a=1$ is the optimum choice.

\section{Conclusion}\label{sec:con}
In this paper, we presented a novel family of linear channel equalization algorithms, based on the logarithmic cost functions, and introduced two specific members of this family, i.e., LCLMA and LCLMS equalizers, as robust adaptive equalizers for underwater acoustic channels. We showed that these algorithms can be implemented in a decision feedback equalization framework, in a similar manner. In the impulse-free environment, the LCLMA algorithm has a similar convergence performance with the famous LMS algorithm, whereas providing a better robustness against impulsive interferences. Furthermore, the LCLMA equalizer combines the fast convergence of the LMS equalizer and the better convergence of the SA equalizer with similar computational complexity. We provided the tracking performance of the LCLMA algorithm both in the impulse-free and impulsive noise environments. Finally, we showed the enhanced performance of the new algorithms in a realistic underwater communication scenario.
\begin{acknowledgements}
This work is in part supported by Turkish Academy of Sciences, Outstanding Researcher Programme and TUBITAK, Contract No:112E161.
\end{acknowledgements}
\bibliographystyle{IEEEbib}
\bibliography{myReferences}   

%



%
%

\end{document}